# Spectrum Allocation in Cognitive Networks

Himanshu Agrawal
CSE Dept.
JIIT-128
NOIDA, India
himanshu.agrawal@jiit.ac.in

*Abstract*—Cognitive Network is a technique which is used to improve the spectrum utilization. Current network scenario is experiencing the huge spectrum scarcity problem due to the fixed assignment policy so in this method great amount of spectrum remain unused. To overcome this limitation the spectrum allocation must be in dynamic manner. In this paper the spectrum allocation is discussed thoroughly. Interference is the most important factor that needs to be considered. It is caused by the environment (noise) or by other radio users. It limits the possibility of spectrum reuse. Channel assignment is one of the techniques used to control interference in the network. There exist a trade-off between network capacity and level of contention. In cognitive radio networks spectrum assignment or spectrum allocation or frequency assignment is used to avoid interference. It is the process of simultaneous selection of operating central frequency and bandwidth.

In doing so, the process of sensing the spectrum becomes very crucial; it must be reliable, accurate and efficient. The accuracy of sensing affects the overall operation of cognitive networks. Accurate results not only lead to higher utilization of the spectrum but also preserve the privacy of primary user. Accuracy of sensing is highly affected by the natural causes like noise, shadowing, fading etc. There are many other challenges as well, like, hardware requirements, hidden node problem, security, sensing frequency and duration, decision fusion etc.

## I. INTRODUCTION

Cognitive Networks [1] are also known as the Next Generation (xG) Wireless Communication Networks [4] as well as Dynamic Spectrum Access Networks (DSANs) [4]. It provides opportunistic [1] access to the spectrum of license bands without affecting the current users. The basic building block of the network is cognitive radio[1]. It is based on the software defined radio (SDR) [2]. The front end of cognitive radio is required to be developed while rest of the part is same like other conventional radios. Cognitive radio is defined as an intelligent communication device that can sense its surrounding environment and can adapt to operate accordingly. It is based on the concept of understanding by building [1] to learn from the environment. The main objective of cognitive radio is reliable communication and efficient utilization of spectrum. It provide the dynamic spectrum access i.e. spectrum sensing (to determine the available spectrum), spectrum management (select the best available channel), spectrum mobility (moving from one spectrum to another when required), spectrum sharing (spectrum is shared by other xG users)[4].All these functionality should cooperate to improve the spectrum efficiency and it requires the spectrum aware communication protocols.

## II. COGNITIVE RADIO

The characteristics of cognitive radio arethat it can change its transmitter parameter according to the environment. The main features of cognitive radio are cognitive capability and reconfigurability [4]. Cognitive capability means the ability to extract the required information from the surrounding environment. It will lead to the selection of best available channel and appropriate operating parameters. Reconfigurability deals with the dynamic programming of cognitive radio based on the environment. It can operate on various ranges of frequencies.The concept of cognitive radio was given by Mitola [5] in 1999. It mainly focuses on the radio knowledge representation language (RKRL) [5]. It is a new language that enhances the flexibility of cognitive radio. The background of cognitive networks is based on the early study of Federal Communication Commission (FCC) in November 2002 about electromagnetic radio spectrum in US [1]. According to that the spectrum utilization of various geographical locations varies from 15% to 85%. It is a very scarce resource. The efficient utilization of this resource is very important.

In cognitive networks there is a lot of computing in terms of language understanding by the radio and processing of the signals etc.Interference temperature is a new standard for the approximation and managementof interference in a radio environment. There is a concept of Interference temperature limit that defines the maximum permissible amount of the interference in the radio environment. Transmission in that band should be avoided if it would increase the noise floor above the interference temperature limit. Estimation of the Interference Temperature, and Detection of Spectrum Holes, are two adaptive, dynamic and correlated functions. To estimate the interference temperature and spectrum holes receiver is required to use a large number of sensors to properly sense the RF environment. So it is very obvious that a reliable strategy is required to detect the spectrum holes.The available spectrum holes depend upon the particular time and location. It requires dynamic spectrum management techniques. It should continuously monitor the spectrum holes.In case of cognitive radio it is essential that it must be transmitter centric because they operate in a decentralized manner like in adhoc networks.

### III. COGNITIVE NETWORKS

There are two portions of next generation network architecture i.e. primary network and xG network [4]. Primary network is referred to the existing network i.e. cellular or TV broadcast networks. They are infrastructure based networks. The components of a primary network are primary user and primary base station. Primary user is also known as licensed user having access to the primary base station. They do not need any modification for sharing with xG users. Primary base station has license to operate in a certain spectrum i.e. base station transceiver system (BTS). They must have some sort of xG protocols to share the spectrum with xG users. xG network can be deployed as an infrastructure based network or as an adhoc network. It accesses the spectrum in an opportunistic manner. It is also known as cognitive radio network or dynamic spectrum access network. The components of xG network are xG user, xG base station and spectrum broker [4]. xG user does not have any license hence they require additional functionality to access the licensed spectrum. They are also known as secondary users, unlicensed users or cognitive users. xG base station provides access to the users without having spectrum license. It has the capability of providing single hop connection to the xG users. Spectrum broker is a network component that facilitates the sharing of the spectrum among different xG users. The network architecture can be defined in licensed as well as unlicensed band. If it operates in a licensed band then the care should be taken to avoid the interference with the existing users (licensed users). The xG user should operate in a licensed band when it is free i.e. there is no licensed user present. On the other hand when it is operated in an unlicensed band then the interference should be avoided among them. Scheduling of spectrum is required to avoid interference.

### IV. SPECTRUM SHARING

The nature of wireless network provides shared access of the resources. The very scarce natural resource is spectrum. The spectrum is shared by the primary users as well as secondary users. The coexistence of these users imposes a lot of challenges in the implementation of cognitive networks. There are some aspects of the spectrum sharing: the architecture of the network, spectrum allocation policy and spectrum access mechanism.

Architecture [8] – Two types of spectrum sharing architectures exist: centralized and distributed. In centralized architecture a central authority is responsible for allocation of the spectrum. The sensing is done individually by the nodes i.e. distributed sensing and then this data is forwarded to the central authority. The allocation map is constructed. This is used to lease the spectrum to the users in a particular location as well as for a particular time interval. In this approach if the centralized authority fails then the whole system collapses. In distributed architecture the allocation is done by the message passing among nodes. This is the process in which all the competing nodes sense the medium and then pass the messages to each other. It degrades the spectrum efficiency. The allocation policy is local or sometimes it is globally followed by all the nodes.

Allocation Policy–The allocation behavior can be cooperative or non cooperative. In cooperative policy the nodes shared the spectrum monitoring data with each other. In this approach a cluster is formed and the decision is taken locally. It is a mid way of centralized and distributed architecture approach. In non-cooperative approach there is no message passing between nodes. Only a single node is considered. Due to low information exchange it consumes more energy in transmission and other control processes. The cooperative approach provides the fairness and improved performance.

Intranetwork Spectrum Sharing- The spectrum is shared among the cognitive users inside the cognitive network. The resources are shared by the nodes locally they does not cause interference to the primary users. There are various challenges in this particular approach.

Internetwork spectrum sharing- The network is shared by the base stations of different cognitive networks to provide adequate quality of service and to fulfill requirement of stations of their individual network. It is the concept in which multiple cognitive networks share the spectrum in same geographic location.

### V. CHALLENGES IN SPECTRUM SHARING

There are various challenges in this spectrum sharing approach: common control channel, dynamic radio range, spectrum unit, location information.

Common Control Channel – In cognitive networks it is not possible to maintain a common control channel because when primary user wants a channel it is required to vacate the space immediately. It can be maintained locally like for clustered scenarios. The common control channel is useful for administrative purpose like spectrum sharing and monitoring.

Dynamic radio range–In CR networks the radio can change its operating frequency therefore in such case the neighbors are dynamic. It means that whenever frequency changes their neighbor also changes. There is interdependency between operating frequency and radio range. It is very crucial issue and it needs to be resolved.

Spectrum unit – The communication channel is considered to be as a spectrum unit. So it is required to be considered in developing algorithms.

Location information – There is an assumption in CR networks that secondary users know the location of primary users as well as their transmission power. But practically this is not always a valid assumption.

## VI. SPECTRUM SENSING

In Cognitive Networks Spectrum Sensing and Spectrum Sharing are the most crucial and critical phenomenon to be studied thoroughly and deeply. Certain issues must be kept in mind like maximize throughput, interference and secondary user's existence. The main issues of cognitive radio are sensing, measurement, learning, power allocation, user's requirement, legal issues, operating constraints etc. The scenario of secondary users is like they should not create any type of interference to the primary users who are legitimate and higher priority to the spectrum access. Cognitive radio is used by the secondary users. For this purpose there is a requirement of an algorithm that can monitor such type of operation in the network. Therefore the spectrum sensing is the most important concept in order to establish such type of cognitive networks.

The report of FCC [10] (Federal Communication Commission) about the spectrum utilization is the basis for the development of Cognitive Radio. This report is also basis for the report of downtown Berkeley that indicates the very low utilization of spectrum especially in low frequency band like 3-6 MHz and it inspire the development of more intelligent way of spectrum usage. The FCC has issued a notice to promote the use of cognitive radio technology. It is known as NPRM (Notice of Proposed Rule Making) to implement the opportunistic spectrum access (OSA) [12]. One of the ways is dynamic spectrum allocation rather than existing static spectrum allocation methods. But it requires the change in the whole communication system as well as infrastructure which are not possible because of huge investments. So it is necessary to developsuch a technique that can adaptively use the unused spectrum and improve the spectrum utilization. This dynamic technique requires certain level of intelligence because it has to identify the unused spectrum from the available whole bandwidth. So there is a requirement of machine learning concept in order to implement this technique efficiently and effectively. The radio must be intelligent enough to understand this requirement of spectrum sensing accurately and efficiently. In the current wireless domain the requirement of the spectrum has increased tremendously. The demand is increasing day by day. So there is always a requirement of the spectrum in the wireless domain. The high frequency bands are still not heavily used so in order to use this spectrum there is a requirement to design the radio which is able to communicate in the high frequency bands. The concept of Cognitive Radio came into existence after the proposal given by Mitola in his thesis. The cognitive radio is very useful and intelligent device in order to improve the spectrum utilization. But there are certain practical limitations in order to develop this radio. Like it must be able to sense the spectrum very quickly and accurately and access it intelligently. It must be adaptive to switch to another channel whenever any legitimate user is trying to access the same frequency. This handoff time must be minimized so that the communication should be unaffected. There are various ways of improving the spectrum sensing capability of the radio. Spectrum Sharing is the Dynamic spectrum access mechanism where different users freely access to some common spectrum under their mutual limited interference. In cognitive network it is very important to maintain the Nash equilibrium i.e. a concept of game theory where the optimal outcome of a game is one where no player has an incentive to deviate from his or her chosen strategy after considering an opponent's choice. Overall, an individual can receive no incremental benefit from changing actions, assuming other players remain constant in their strategies. A game may have multiple Nash equilibriums or none at all.

Spectrum sensing is defined as a process to obtain the information related with the spectrum like awareness of spectrum usage and existence of primary users. It is the process of measuring the radio frequency energy in the spectrum. There are four different domains in which sensing are required to be done i.e. space, time, frequency and code. It is required to consider the modulation, carrier frequency and bandwidth of the spectrum. In cognitive networks the computational complexity is very high and it requires efficient signal analysis techniques as well. The operation of secondary users depends upon the opportunistic access of the spectrum. It means that it requires sensing and exploiting the spectrum space efficiently. The identification of opportunity involves the existence of band of frequency that are not used by the primary user in particular time, frequency at particular location i.e. space. So there are three things space, time and frequency. There is one more dimension i.e. code that need to be consider but traditional algorithm does not take this into account. There are lots of challenges in the identification of the opportunity in the spectrum. Spectrum sensing is the process that should include the identification of spectrum holes.

## VII. CHALLENGES

Hardware Requirements- Spectrum sensing in real time is very crucial. It requires very high performance hardware because very complex computations take place in sensing algorithms. For sensing it requires very high sampling, analog to digital conversion, high speed signal processing, channel estimation, power control etc. It requires efficient receivers to cover very wide range of frequencies. Delay is very important aspect while sensing the spectrum therefore it should be minimized. There are two ways to perform sensing single radio and dual radio. In single radio a specific time slot is devoted for the purpose of sensing which is the wastage of precious natural resource i.e. spectrum. But this approach is easy to implement and low cost as compare to dual radio. In dual radio one radio is used for communication and other one is used for sensing. In this approach spectrum efficiency is high as compare to previous approach. In single radio very short time duration is allocated for sensing therefore accuracy is suspected. In dual radio power consumption and cost are increased as well.

Hidden Licensed User Problem- It is due to the many factors like shadowing, multipath fading, it is observed by the unlicensed users while searching for the licensed users in the spectrum. It is similar to the hidden node problem in wireless networks. It causes interference to the licensed user because secondary user is not able to detect the location of primary user but when both transmit at the same time interference occur. Cooperative sensing is one of the ways to handle this particular problem.

Sensing Duration and Frequency- In cognitive networks the primary users communicate in their licensed band and secondary users communicate in the band when it is not used by the primary users. But primary users can start communication whenever they wanted to do so it is essential for secondary users to sense the presence of the primary users as soon as possible and switch to another free band. This process requires the frequent sensing of the spectrum for primary users and it is essential to avoid the interference with primary users. It is a great challenge for sensing algorithms. There is a tradeoff between sensing accuracy and its duration. Sensing frequency is very important parameter that is to chosen very carefully. It depends upon the channel characteristics, behavior of the network and primary users. If the state of licensed users changes frequently, then it will be a great challenge for sensing algorithms because the sensing frequency is also required to be increased. There are some other parameters as well that require to be considered like, sensing period, time required for channel detection and channel movement. There are many other timing related parameters as well. The sensing frequency affects the performance of the secondary users very badly throughput is degraded due to these sensing processes. So, it is required to get the optimum value of sensing duration and frequency. Sensing duration can be minimized by scanning only changing part of the space compare to whole spectrum. It is not possible for secondary users to transmit the data and sense the space at the same time on the same channel. It is required to interrupt the transmission for sensing that result in the degradation of spectrum efficiency.

Security– A selfish cognitive radio user or secondary user can impersonate as a primary user by changing its air interface parameters. It misleads the scanning of space performed by the legitimate secondary users. This is known as primary user emulation attack. To counter such an attack public key encryption algorithms are used and it is essential that primary users send an encrypted value that can be generated by their private key. It acts as their signature. It can be used only with digital modulation techniques and it requires that secondary users should be capable of demodulating the signals of primary users.

Interference Temperature Measurement–It is one of the most important aspects of cognitive networks. There is no interaction between primary and secondary network users. There is requirement of new techniques to measure this aspect.

Multiuser Environment – It is very difficult to sense the spectrum in a multiuser environment because there exists multiple primary and secondary users, therefore it is pretty difficult to identify the spectrum holes and the estimate of interference temperature.

Spectrum Efficient Sensing- The process of sensing cannot be performed at the time of transmission by the cognitive radio. For sensing the transmission must be stopped hence it degrades the spectrum efficiency. Therefore it requires a balance between sensing accuracy and efficiency. Although the sensing time directly affects the sensing accuracy, therefore such sensing algorithms are required that can provide sufficient accuracy with minimized sensing duration.

## VIII. SENSING METHODS

There are mainly two ways of spectrum sensing. First is energy detection and second is cyclostationarity. It is the property of digital modulated signal. This is an advance process that is capable of signal classification and ability to identify co channel interference. There are three types of spectrum holes white, grey and black. White denote the absence of any signal and at the same time it is best suited for transmission, and grey denote the presence of interference so it is requires to identify the presence of users signal further. And black denote the presence of users signal, cannot be used by other user at the same time. These two approaches can be applied only in the case of white spaces only. Therefore their application is limited. In wireless network reliable communication is must and in order to increase the utilization of spectrum, it is require having a reliable method of spectrum sensing. It should be capable of high spectral resolution and computationally feasible in real time.

There are many algorithms for the purpose of spectrum sensing, some of them are as follows.

- Matched Filtering
- Energy Detector
- Spectral Correlation
- Radio Identification Based Sensing
- Waveform Based Sensing
- Multi-Dimensional Spectrum Sensing
- Internal Sensing
- External Sensing
- Cooperative Sensing

Multi-Dimensional Spectrum Sensing [2] – Spectrum Sensing can be carried out in different dimensions like time, space and frequency. Apart from that there are other dimensions are also available like code, angle etc. the traditional algorithm are not able to dealt with these dimensions. They can deal with traditional dimensions only. Now a new concept of multi antenna radio is also came recently in cognitive networks. It enables the beam forming in the network. By doing so it is possible to access the spectrum by more than one user at the same time and in the same frequency band. There are some advantages and disadvantages of using multi antenna system in the network like single radio is simple, low cost but lower spectrum efficiency and poor sensing accuracy. On the other hand double radio is costly, quite complex and higher power consumption but its spectrum efficiency is high and sensing accuracy is also better. There are multiple dimensions like frequency, time, space, code and angle in which it can be sensed for the hole.

Energy Detector Based Sensing [5] – This process is also known as radiometry. It is the most common methods of spectrum sensing. It is the easiest method of sensing having low computational and implementation complexities. In this method the output of the signal is compared with the threshold that depends on the noise floor. In this method receiver does not know the primary users signal information.

Limitations – It is not able to differentiate between interference from primary user and noise. One more thing is its dependency on the threshold. It means it is complex to decide the proper threshold value. Its performance degraded under the low signal to noise ratio and it is not able to detect the spread spectrum signals. There is a concept of probability of detection and probability of false alarm. It is desired to have high $P_d$ and lower $P_f$. It will result in the high utilization of spectrum otherwise lower utilization will be resulted. The threshold is decided on the basis of the
noise variance. Therefore small error in the estimation of the noise variance will result in the significance degradation in the performance of the network.

Web Form Based Sensing [6] – In this method signal is sensed through the matching by itself. It means that the copy of the signal should be available at the receiver end and then it will be matched by the receiver. This method is better than previous method because it is reliable and having less convergence time.

Limitations – This method can only be applied in the case of signal with known patterns. This method requires short measurement time but it is vulnerable to synchronization errors.

Cyclostationarity Based Sensing [9] – In this method cyclic correlation function is used to detect the presence of the signal in the spectrum. It is able to differentiate the noise and primary user signal. Cyclostationary features are caused by the repetitions in the signal.

Radio Identification Based Sensing [11] – The identification of technology, used by the primary users can be very useful to detect the signal efficiently. This complete knowledge about the spectrum will enable the cognitive radio to generate the accurate result regarding the presence of primary user. In this technique the features of the signal are extracted and then they are used for the classification. Features like energy and its distribution across the signal spectrum, channel bandwidth etc are very important and distinguished features of the signal.

Matched Filtering [11] – This method is most suitable when the transmitted signal is known. The advantage of this method is its short response time to calculate the probability of false alarm and it can work in low SNR as well. But it requires the demodulated received signals so it requires the knowledge of the primary users signal features like bandwidth, frequency, type of modulation, frame format etc. the limitation of this technique is that the complexity of the receiver is very high and its power consumption is also very high.

Comparison of Various Sensing Techniques – Waveform based sensing is more robust than energy detection method and cyclostationary method. But it requires that the primary users signal characteristics should be available at the receiver and it should transmit the known patterns. There are two assumptions in the energy detection method that must be hold like noise should be stationary and its variance must be known. Otherwise the cyclostationary method is most suitable. But it also affect in the case of fading.

Energy Detector Based Sensing - In this method the received signal is detected by comparing the energy detector with a threshold that depends on the noise floor [18]. It is not able to differentiate between the primary user's signal from interference and white Gaussian noise. Its performance is very poor when signal-to-noise ratio (SNR) values are very low.

Cyclostationarity - Based Sensing - In this method primary user transmission are detected and it can distinguish noise from primary user's signals [18].

Matched-Filtering - This is an optimum method for spectrum sensing only when the transmitted signal is known [18]. It requires perfect and pre knowledge of the signal transmitted by the primary users.

Cooperative Sensing [3] – Cooperation can solve the problem of hidden node. It reduces the probability of false alarm. It solves the issue of shadowing, noise, fading etc. considerably. Challenges to this type of algorithm are its complexity and it also requires the information sharing algorithms. In this method the control channel can be implemented by using unlicensed band such as ISM. It is used for sharing the channel allocation information and spectrum sensing results.

Centralized Sensing [15] – In this a centralized unit will collect the information from different cognitive devices and then find the availability of spectrum holes and broadcast this information. It uses the large bandwidth for this control information if the number of users is very high. To overcome this problem the censoring can be applied it means only those users are allowed to submit their information who are reliable.

Distributed Sensing [15] – In this technique the cognitive users share their information and take their own decision locally that which part of spectrum can be used by them. It reduces the cost of control channel as there is no need of broadcasting huge information.

External sensing [15] – In this approach an external agency is used to sense the spectrum and it will broadcast the channel occupancy information to cognitive users. It is a better approach as compare to internal sensing because it will solve the problem of hidden nodes and the problem of shadowing and fading. And in internal sensing the nodes have to sense the spectrum so spectrum efficiency is degraded while in external sensing, spectrum efficiency is improved because it is done by the external agency. The sensing units are not required to be run on batteries and it is not required to be mobile. So it solves the problem of power consumption in internal sensing.

## IX. SPECTRUM ALLOCATION

The aim of this approach is to limit the interference so that the capacity and performance of the network can be improved. This process is related with spectrum sensing. Spectrum sensing is responsible for finding the available frequency band then spectrum decision or spectrum assignment is based on the parameters like fairness, quality of service requirement, throughput, spectrum efficiency etc. and the constraint is, interference to the primary user as well as secondary users must be avoided. The task of spectrum assignment is carried out by selecting the central frequency and bandwidth (according to requirement) simultaneously.

The process of spectrum assignment is divided into three parts. First is define the objective function that is criteria to solve this problem then the suitable modeling is chosen and the third step is the selection of technique that will simplify the SA problem.

Criteria – There are different criteria for assigning the spectrum to the secondary users. Different criteria are as follows.

- Interference/Power – Interference is the main criteria in order to decide the spectrum allocation algorithm. In the past efforts there are three different cases of interference consideration. First is interference experienced by the single secondary user only, second is interference experienced by all secondary users and third is interference experienced by all users including primary and secondary users.

- Interference Temperature Limit - It is the amount of interference at the receiver end. It is the ratio of power at the receiver end to RF bandwidth and boltzman constant. To limit the interference temperature the approach of power control is used. There is a tradeoff i.e. decreasing the power will result in the decrease in the interference but it will result in low SNR.

- Maximize Spectrum Utilization – This is the most basic criteria to design the spectrum allocation algorithm. In this it is require to maximize the number of channels to secondary user or number of secondary users are require to be maximize.

- Throughput – It should be maximized and basic criteria for selection of algorithm of spectrum assignment. There are some constraints like quality of service requirement, SINR, maximum permissible transmission power and link capacity.

- Fairness – To achieve fairness it is require having a central control unit because maximization of throughput does not guarantee the fairness in the network i.e. maximization of each and every secondary user's throughput.

- Delay – It is one of the qualities of service criteria. There are two types of delay experience by the network, switching delay and end to end delay.

Spectrum Allocation Procedures – There are different procedures of spectrum allocation. Like centralized, distributed, clustered and inclusion of primary user.

- Centralized or Distributed Approach – In centralized approach a server is used to collect the spectrum related information of the network and it decides the spectrum allocation based on some predefined criteria. It preserves the fairness of the network and maintains the required quality of service. It depends on the server for the functioning of the network. While in later case the approach is quick and based on cooperation among the nodes but it is not able to preserve the fairness.

- Existence of Primary User – In the literature there are two approaches i.e. consider the presence of primary user or not. In the simplest form the primary users are not considered and it is assumed that fixed set of channels are available for secondary users. Primary users are just uses for the sake of limiting the number of total available channels. While the later work do consider the existence of primary users and then try to find out the way of allocation the spectrum such that it do not cause the excessive interference to the primary user.

- Cluster based – In this approach network is partitioned into clusters and the cluster head is communicating with the base station after collecting the data from cluster members. It is energy efficient and need less number of transmissions of control information. In another implementation cluster head itself can take the decision of spectrum allocation after cooperating with other cluster heads.

## X. CONCLUSION

In Cognitive Networks the main goal is to improve the Spectrum Utilization. There are three steps for doing this- (1) Spectrum sensing, (2) spectrum decision or spectrum allocation and (3) spectrum handoff or mobility. In this paper, procedures, techniques and criterion, for spectrum sensing and then allocation has been studied and it is understood that better allocation can drastically improves spectrum utilization.

There are various challenges in order to adopt this method.
- Complete network architecture that can handle the heterogeneities at various levels.
- Environment effects should be considered explicitly.
- Upper layer challenges like congestion, routing, frame errors etc.
- The memory and processing constraints of mobile terminals is required to be considered.
- It is required to develop an architecture that can accommodate the various environments like WLAN, 3G cellular, satellite network, ISM band etc. it facilitate the global roaming of mobile user. It enables the service providers to serve "Anywhere and Anytime".

REFERENCES

[1]. Simon Haykin, "Cognitive Radio: Brain-Empowered Wireless Communications", IEEE Journal on Selected Areas in Communications, Volume 23, Number 2, February 2005, Pages 201-220.
[2]. Friedrich K. Jondral, "Software-Defined Radio—Basics and Evolution to Cognitive Radio", EURASIP Journal on Wireless Communications and Networking, 2005, Pages 275–283.
[3]. Ozgur B. Akan and Ian F. Akyildiz, "ATL: An Adaptive Transport Layer Suite for Next – Generation Wireless Internet", IEEE Journal on Selected Areas in Communications, Volume 22, Number 5, June 2004, Pages 802-817.
[4]. Ian F. Akyildiz, Won-Yeol Lee, Mehmet C. Vuran and Shantidev Mohanty, "Next Generation/Dynamic Spectrum Access/Cognitive Radio Wireless Networks: A Survey", Computer Networks 50, 2006, Pages 2127–2159.
[5]. J. Mitola et al., "Cognitive radio: Making software radios more personal," IEEE Personnel Communications, Volume 6, Number 4, August 1999, Pages 13–18.
[6]. M. Turner, "JTRS application in cognitive technology," presented at the Conference of Cognitive Radios, Technology Training Corporation, Las Vegas, NV, March 15–16, 2004.
[7]. Raul Et kin, Abhay Parekh, and David Tse, "Spectrum Sharing for Unlicensed Bands", IEEE Journal on Selected Areas in Communications, Volume 25, Number 3, April 2007, Pages 517-528.
[8]. Kamrul Hakim, Sudharman K. Jayaweera, Georges El-howayek, and Carlos Mosquera, "Efficient Dynamic Spectrum Sharing in Cognitive Radio Networks: Centralized Dynamic Spectrum Leasing (C-DSL)", IEEE Transactions on Wireless Communications, Vol. 9, No. 9, September 2010, Pages 2956–2967.
[9]. Rocco Di Taranto, PetarPopovski, Osvaldo Simeone, and Hiroyuki Yomo, "Efficient Spectrum Leasing via Randomized Silencing of Secondary Users", IEEE Transactions on Wireless Communications, Vol. 9, No. 12, December 2010, Pages 3739–3749.
[10]. Sudharman K. Jayaweera, and Tianming Li, "Dynamic Spectrum Leasing in Cognitive Radio Networks via Primary - Secondary User Power Control Games", IEEE Transactions on Wireless Communications, Vol. 8, No. 6, June 2009, Pages 3300-3310.
[11]. Xin Wang, "Joint Sensing-Channel Selection and Power Control for Cognitive Radios", IEEE Transactions on Wireless Communications, Vol. 10, No. 3, March 2011, Pages 958-967.
[12]. Tevfik Yucek and Huseyin Arslan, "A Survey of Spectrum Sensing Algorithm for Cognitive Radio Applications," IEEE Communications Surveys & Tutorials, Vol. 11, No. 1, 2009, Pages 116-130.
[13]. Ying-Chang Liang, Yonghong Zeng, Edward C. Y. Peh, and Anh Tuan Hoang, "Sensing – Throughput Tradeoff for Cognitive Radio Networks", IEEE Transactions on Wireless Communications, Vol. 7, No. 4, April 2008, Pages 1326–1337.
[14]. Igor Stanojev, Osvaldo Simeone, Umberto Spagnolini, Yeheskel Bar-Ness, and Raymond L. Pickholtz, "Cooperative ARQ via Auction-Based Spectrum Leasing", IEEE Transactions on Communications, Vol. 58, No. 6, June 2010, Pages 1843–1856.
[15]. Danijela Cabric, Shridhar Mubaraq Mishra, and Robert W. Brodersen, "Implementation Issues in Spectrum Sensing for Cognitive Radios", Berkeley Wireless Research Center, University of California, Berkeley, IEEE, 2004, Pages 772-776.
[16]. Hyoung-Jin Lim, Dae-Young Seol, and Gi-Hong Im, "Resource Allocation for Mitigating the Effect of Sensing Errors in Cognitive Radio Networks", IEEE Communications Letters, Vol. 14, No. 12, December 2010, Pages 1119–1121.
[17]. Sina Maleki, Ashish Pandharipande, and Geert Leus, "Energy - Efficient Distributed Spectrum Sensing for Cognitive Sensor Networks", IEEE Sensors Journal, Vol. 11, No. 3, March 2011, Pages 565-573.
[18]. Simon Haykin, David J. Thomson, and Jeffrey H. Reed, "Spectrum Sensing for Cognitive Radio", Proceedings of the IEEE, Vol.97, No. 5, May 2009, Pages 849-877.
[19]. Yuan Wu and Danny H. K. Tsang, "Joint Rate – and – Power Allocation for Multi- channel Spectrum Sharing Networks with Balanced QoS Provisioning and Power Saving", Springer Science Mobile Network Applications, Issue 14, Published online 24 January 2009, Pages 198-209.